\documentclass[pdflatex,sn-mathphys-num]{sn-jnl}

\usepackage{graphicx}%
\usepackage{multirow}%
\usepackage{amsmath,amssymb,amsfonts}%
\usepackage{fullpage}
\usepackage{ragged2e}
\usepackage[font=small,labelfont=bf,justification=justified]{caption}
\usepackage{soul}

\begin{document}

\title{The Galactic Pizza: Flat Rotation Curves in the Context of Cosmological Time-Energy Coupling}

\author{\fnm{Artur} \sur{Novais}}\email{arturnovais@gmail.com}

\author{\fnm{A.L.B.} \sur{Ribeiro}}\email{albr@uesc.com.br}

\affil{\orgdiv{Laboratório de Astrofísica Teórica e Observacional}, \orgname{Universidade Estadual de Santa Cruz}, \orgaddress{\street{Rodovia Jorge Amado}, \city{Ilheus}, \postcode{45662-900}, \state{Bahia}, \country{Brazil}}}


\abstract{
    The phenomenon of augmented gravity on the scale of galaxies, conventionally attributed to dark matter halos, is shown to possibly result from the incremental growth of galactic masses and radii over time. This approach elucidates the cosmological origins of the acceleration scale \(a_0\approx cH_0/2\pi\approx10^{-10}\){ms}$^{-2}$ at which galaxy rotation curves deviate from Keplerian behavior, with no need for new particles or modifications to the laws of gravity, i.e., it constitutes a new explanatory path beyond Cold Dark Matter (CDM) and Modified Newtonian Dynamics (MOND). Once one formally equates the energy density of the universe to the critical value (\(\rho=\rho_c\)) and the cosmic age to the reciprocal of the Hubble parameter (\(t=H^{-1}\)), independently of the epoch of observation, the result is the Zero-Energy condition for the cosmic fluid's equation of state, with key repercussions for the study of dark energy since the observables can be explained in the absence of a cosmological constant. Furthermore, this mass-energy evolution framework is able to reconcile the success of CDM models in describing structure assembly at \(z\lesssim6\) with the unexpected discovery of massive objects at \(z\gtrsim10\). Models that feature a strong coupling between cosmic time and energy are favored by this analysis.}

\keywords{Gravitation; Cold Dark Matter; Dark Energy; Spiral Galaxies; High-redshift Galaxies; Modified Newtonian Dynamics}



\maketitle

\section{Introduction}

The prototypical case of augmented gravity for which one invokes dark matter is the flattening of galaxy rotation curves, originally observed by Rubin et al. \cite{Rubin1970RotationoftheAndromedaNebula,Rubin1983Rotationalproperties}. Nevertheless, this is far from an isolated case: after decades of research, the dark matter hypothesis has gained support from multiple fronts, e.g., the Cosmic Microwave Background (CMB) power spectrum \citep{PlanckCollaboration}, cluster dynamics and mass distribution \citep{Paraficz2016}, primordial nucleosynthesis \citep{Cyburt2016}, and structure formation modeling \citep{Illustris2014}. On the other hand, there exist a number of tensions and conflicts facing the favored Cold Dark Matter (CDM) paradigm (for a summary on galactic scales, see \cite{deMartino2020}). The most significant challenge is the accumulated null results in the attempts at detection, both on astronomical and laboratory scales, as reviewed \mbox{in \cite{Bergström2007,Misiaszek2024}.}

Meanwhile, observations of the cosmos bring evidence for two remarkable realizations: (i) the universe is flat, meaning that the total energy density is equivalent to the critical value (\(\rho_0=\rho_c\)) \citep{Efstathiou2020TheEvidence}; (ii) the current age of the universe is generally consistent with the reciprocal of the Hubble constant (\(t_0\approx {H_0}^{-1}\)) \citep{Kutschera2007Coincidence}. If one considers the possibility that such results are not coincidences of the present epoch, assuming instead that they would persist whether the observation is performed in the present, billions of years in the past, or billions of years in the future, an interesting description of cosmic evolution emerges. The overall picture of a universe that starts maximally hot and dense, expands and cools over time, undergoes thermal conditions that allow for nucleosynthesis, releases the CMB, and enables the gradual formation of ever larger cosmic structures remains the same as the standard \(\Lambda\)CDM model. However, in this new description, the composite energy density of the universe scales exactly with the inverse square of the cosmic age (\(\rho \propto t^{-2}\)) with no transitory phases, resulting in a matter-energy content that is linearly proportional to time \((U\propto t)\). We call this property “Cosmological Time-Energy Coupling'' (CTEC).

Models that meet the CTEC conditions already exist as emerging candidates to resolve current tensions facing the standard model. Melia and Shevchuk
\cite{Melia2012Rh=ct}, for example, used the Friedmann-Lemaître-Robertson-Walker (FLRW) metric plus the “Zero Active Mass'' constraint \mbox{(\(\rho+3p=0\))}, where \(p\) is the pressure, to develop the \(R_h=ct\) universe. More recently, Novais and Ribeiro \cite{Novais2024} put forward the ZEUS model (standing for Zero-Energy Unified Substratum), a cosmological description based upon the observed spatial flatness and Lorentzian time dilation, to liberate cosmology from the “cosmic reference frame'' and better constrain the evolution of the CMB temperature. These CTEC-type models are shown to alleviate the early galaxy formation problem (pointed out, e.g., in \cite{Steinhardt2016TheImpossibly,Melia2014PrematureFormation,Robertson2023IntenseStarFormingGalaxies,McGaugh_2024}), and we elaborate on this graphically in Section \ref{sec6}. On the other hand, these models need yet to reproduce the \(\Lambda\)CDM success in detailed calculations involving the CMB fluctuations and structure evolution modeling. These developments will be scrutinized in future works.

Another relevant theory to be discussed within our scope is that of Scale Invariant Vacuum (SIV) (see \cite{universe6030046} and the references therein for a complete review). SIV is founded on the Integrable Weyl Geometry and follows the hypothesis that the macroscopic empty space is scale-invariant. One of its properties is that the inertial mass of a particle is not a constant but a function of the conformal scale factor \(\lambda\), ultimately growing over cosmic time. Hence, this theory exhibits a type of time-mass coupling that generates common ground with the models cited above. However, both theoretical and observable differences also exist, and a subset of those are herein described so that the models can be compared on experimental grounds.

As a result of studying the behavior of cosmic energy over time, the theoretical advancement herein reported is closely related to the dark entities, starting with the CDM paradigm and the flattening of spiral galaxies' rotation curves. It is shown that a continuous matter-energy growth of the expanding cosmic fluid gives rise to the acceleration scale \(a_0\approx cH_0/2\pi\) at which galactic rotation curves deviate from Keplerian decline. The relevant novelty, therefore, is that one need not alter the known laws of gravity nor invoke an extension to the Standard Model of particle physics, building a third path to account for the phenomena of augmented gravity that extends beyond Modified Newtonian Dynamics (MOND) \citep{Milgrom1983AModification} and the Weakly Interacting Massive Particles \mbox{(WIMPs) \citep{Roszkowski_2018}}.

In summary, this new approach shares the positive results that MOND demonstrates by featuring a critical acceleration scale, e.g., a solution to the galaxy rotation \mbox{problem \citep{Sanders2002}}, the Baryonic Tully-Fisher Relation (BTFR) \citep{TullyFisher1977}, the Radial Acceleration Relation (RAR) \citep{Lelli_2017}, indefinitely flat circular velocities \citep{Mistele_2024}, the dissolution of halo dynamics \mbox{inconsistencies \citep{deBlok2010,Kroupa2015}}, and the solution to various tensions with satellite galaxies \citep{Boylan2012,Kaplinghat:2019svz} while avoiding the setbacks of attempting to modify gravity in specific regimes. Since \(a_0\) emerges from a physical phenomenon, as opposed to a modified gravity framework, it does not run into tensions with conservation laws \citep{Gentile2011THINGS} and does not cause any conflicts with the recent studies on wide binary stars based on the GAIA mission data \citep{Banik2024}.

\section{Newtonian Gravity with Variable Mass and Radius} \label{sec:Newtoninan}

At low accelerations, the field equations of General Relativity reduce to Newtonian dynamics. It is also in this regime that galactic rotation curves deviate from Keplerian behavior. Therefore, the Newtonian language is appropriate for the analysis of augmented gravity under the condition of marginal changes in mass \mbox{and distance}.

On the periphery of spiral galaxies, where the bulk mass sourcing the gravitational pull is no longer a function of distance, the orbital velocities are expected to decrease with the square root of the separation from the galactic center. This Keplerian behavior can be directly extracted from the Newtonian equations for centripetal and \mbox{gravitational accelerations:}

\begin{equation}
a_{cp}=\frac{v_{Kep}^2}{r}=\frac{GM}{r^2} ,\ \ \  v_{Kep}=\sqrt\frac{GM}{r}
\end{equation}

However, the observed rotation curves reveal a flattening of the orbital velocities after a characteristic acceleration scale \(a_0\). The basic idea of the MOND approach proposed by Milgrom \cite{Milgrom1983AModification} is to postulate the presence of this transitional scale in the equations of motion ab initio. By running this exercise, Milgrom identified a potentially relevant aspect: the apparent numerical coincidence between the measured value of \(a_0\) and the combination of fundamental cosmological parameters:

\begin{equation}
a_0\approx\frac{c}{2\pi}H_0\approx 10^{-10}~\text{m/s}^2
\end{equation}

Although the hint is acknowledged, no explanation has been provided within the MOND context yet \citep{Milgrom2015MONDTheory}.

\subsection{Time-Dependent Gravitational Acceleration from Dimensional Analysis}

We now demonstrate that Newtonian equations in their own domain can give rise to an acceleration scale that is a function of cosmic time. As a first step, a straightforward dimensional analysis can be employed. Considering that the Newtonian gravitational constant is indeed fixed \citep{Sakr_2022}, one can assess the theoretical consequences of disk mass and radius growth over cosmic time by expressing them as temporal functions:

\begin{equation}
M_d=At^{\alpha}, \ \ \ r_d=Bt^{\beta}
\end{equation}

\begin{equation}
a=\frac{GAt^{\alpha}}{B^2t^{2\beta}}=\frac{GA}{ B^2}t^{\alpha-2\beta}
\end{equation}

Note that the acceleration scale \(a_0\) can also be written as a temporal function once one assumes a cosmological framework in which the Hubble parameter corresponds to the reciprocal of cosmic time:

\begin{equation}
H= t^{-1}
\end{equation}

When writing explicitly for the present epoch

\begin{equation}
H_0= t_0^{-1}
\end{equation}
and comparing the two resulting expressions for a present acceleration term

\begin{equation}
a_0 \approx \frac{c}{2\pi}t_0^{-1}, \ \ \  a_0=\frac{GA}{ B^2}t_0^{\alpha-2\beta},
\end{equation}
it becomes clear that a connection between Newtonian acceleration and the critical scale identified by Milgrom can be established if the following condition is met:

\begin{equation}
\alpha-2\beta=-1
\end{equation}

The simplest case of a first-order approximation for both mass and radius (\(\alpha=\beta=1\)) obeys this constraint. Moreover, as evidenced next, it yields the transition from an inverse square law (\(a_{cp}\propto r^{-2}\)) to a simple inverse law (\(a_{cp}\propto r^{-1}\)), in agreement with the flat rotation curves and the RAR observations.
So, to conclude this theoretical exercise, \(A\) and \(B\) can be written in a more explicit manner as \(\dot M_d\) and \(\dot r_d\), respectively,

\begin{equation}
M_d=\dot M_dt, \ \ \ r_d=\dot r_dt
\end{equation}
yielding a critical acceleration that is equal to

\begin{equation}
a_c(t)=\frac{G\dot M_dt}{\dot r_d^2t^2}=\frac{G\dot M_d}{\dot r_d^2}\frac{1}{t}=\frac{G\dot M_d}{\dot r_d^2}H(t)
\end{equation}
and for the present time,

\begin{equation}
a_0=\frac{G\dot M_d}{\dot r_d^2}H_0
\end{equation}

Thus, from this simple dimensional analysis, a Newtonian acceleration that is a function of the Hubble constant was obtained. As a final step of this exercise, one can even relate the rates of mass and radius growth by using Equations (2) and (11):

\begin{equation}
\dot M_d \approx \frac{c\dot r_d^2}{2\pi G}
\end{equation}





Naturally, a strictly linear growth is an oversimplification of the structure formation history. For a practical application of this theoretical dimensional analysis, one needs to consider the multitude of intertwined astrophysical processes involved in galactic evolution and the varying scales over which they operate \citep{vandenBosch2002,Kassin2012}. Next, we take the first step to reconcile a more realistic function of galactic growth with this dimensional analysis.

\subsection{Mass and Radius in Galaxy Formation Models}\label{sec2.2}

The growth of galactic masses and radii is influenced by several factors, including the accretion of gas from the intergalactic medium (IGM), mergers with other galaxies, and internal processes, such as star formation, feedback from supernovae, and active galactic nuclei. Each of these components operates over different timescales and spatial ranges, introducing significant variability to how galaxies evolve. Furthermore, the environment in which a galaxy resides---whether isolated or within a dense cluster---can drastically affect its growth trajectory through mechanisms like tidal interactions or ram-pressure stripping~\citep{Hopkins2012,Hester_2006}.

The recent results produced by the James Webb Space Telescope (JWST) have stirred even further discussions on this complex subject matter \citep{Robertson2023IntenseStarFormingGalaxies,McGaugh_2024}, building upon the debates initiated by the Hubble Space Telescope (HST) data \citep{Steinhardt2016TheImpossibly,Melia2014PrematureFormation,Melia2015Supermassive}. The discovery of highly luminous objects at \(z\gtrsim 10\) challenges the hierarchical formation picture characterized by the gradual mergers of smaller progenitors, pointing, instead, to a rapid initial burst of mass assembly---and its conversion into stars---followed by quiescent behavior, i.e., a monolithic process \citep{McGaugh_2024}.

In statistical terms, without going into the details of the complex astrophysical machinery for each particular galaxy, one can propose temporal growth functions that reconcile the rapid evolution observed in early times with the emergence of the acceleration scale \(a_0\) at later times:

\begin{equation}
M=M_f(1-e^{-t/\tau_1})+\dot M_dt, \ \ \ r=r_f(1-e^{-t/\tau_2})+\dot r_dt
\end{equation}

In Section \ref{sec6}, we show that this type of evolution function can successfully reproduce our current structure formation knowledge---based on simulations and data---both for low and high redshifts.

The first terms are characteristic prescriptions for the initial mass and radius buildup. As soon as \(t>\tau_{1,2}\), such terms rapidly approach the values \(M_f\) and \(r_f\). For the strictly monolithic picture, this would mean the end of active evolution, but here, we add a second term describing incremental growth, which is secondary during the initial spurt but will dominate the subsequent evolution (see the first panel of Figure 2 for a graphic representation). Additionally, note that we use the same type of function for radius and mass growth, but they do not necessarily evolve at the same time frame (it can take longer, for example, for the radius to grow once matter has clumped), hence the different labels \(\tau_1\) and \(\tau_2\). However, such distinction does not affect our following analysis provided that both are sufficiently smaller than the Hubble time (\(\tau_{1,2}<<H_0^{-1}\)). Likewise, we consider here that the time it takes for the growth of a gravitationally bound system to begin after the big bang is sufficiently small (\(t_i<<H_0^{-1}\)) so that, instead of writing \((t-t_i)\), we abbreviate it, for now, to simply \(t\).

During most parts of galactic evolution, with the only exception being the brief period of growth spurt at the beginning, the mass and radius are given by

\begin{equation}
M=M_f+\dot M_dt, \ \ \ r=r_f+\dot r_dt
\end{equation}

Equipped with these expressions, one can study the effects of changing mass and radius, characterized by \(\dot M_d\) and \(\dot r_d\), on centripetal accelerations and orbital velocities.

By conveniently rearranging the terms through polynomial division, as follows,

\begin{equation}
a_{cp}=G\left[\frac{M}{r}\right]\frac{1}{r}=G\left[\frac{\dot M_dt +M_f}{\dot r_dt+r_f}\right]\frac{1}{r}=G\left[\frac{\dot M_d}{\dot r_d}+\left(\frac{M_f}{r_f}-\frac{\dot M_d}{\dot r_d}\right) \frac{r_f}{r}\right]\frac{1}{r},
\end{equation}
one can show that the inverse square law (\(a_{cp}\propto r^{-2}\)) turns into a simple inverse relation (\(a_{cp}\propto r^{-1}\)) if one of the following conditions is met: (i) the ratio \(M_f/r_f\) is equal to \(\dot M_d/\dot r_d\), which is a reasonable condition for the formation and evolution of the galaxy since the same environmental aspects that generated the initial growth spurt's mass-to-radius ratio will dictate the behavior of the linear terms; (ii) enough time has elapsed so that \(r>>r_f\).

If one of these conditions is true---or an approximate combination of the two---\(M/r\) approaches the constant \(\dot M_d/\dot r_d\) in the outer regions of the spiral. Consequently, the circular velocities become indefinitely flat (\(v_f\) independent of \(r\)), which conforms with the observations reported by Mistele et al. \cite{Mistele_2024}.

\begin{equation}
a_{cp}=G\left(\frac{\dot M_d}{\dot r_d}\right)\frac{1}{r}=\frac{v_f^2}{r}, \ \ \ \therefore v_f=\sqrt{\frac{G\dot M_d}{\dot r_d}}
\end{equation}

Furthermore, once the late-time condition is met (\(\dot r_dt_0>>r_f\)), the centripetal acceleration turns into the very \(a_0\) we obtained via dimensional analysis:

\begin{equation}
a_{cp}=G\left(\frac{\dot M_d}{\dot r_d}\right)\frac{1}{r}=G\left(\frac{\dot M_d}{\dot r_d}\right)\frac{1}{\dot r_dt_0}
\end{equation}

\begin{equation}
\therefore a_{cp}\approx\frac{G\dot M_d}{\dot r_d^2}\frac{1}{t_0}\approx\frac{G\dot M_d}{\dot r_d^2}H_0
\end{equation}

Thus, the ubiquitous acceleration scale extracted from the observation of galaxy rotation curves and its relation to the Hubble constant can be explained:

\begin{equation}
a_{cp}\approx a_0, \ \ \  \frac{G\dot M_d}{\dot r_d^2}H_0 \approx \frac{c}{2\pi}H_0
\end{equation}

Here, it is worth emphasizing how this physical description differs from that of dark matter. It is not the extra mass given by the minor rate of change \(\dot M_d\) alone that is responsible for the augmented gravity; to achieve that solely based on mass, a major wealth of extra matter would be required, as suggested by the massive CDM halos \citep{Navarro1996}. Rather, it is the combined time-evolution of both \(M_d\) and \(r_d\) that naturally produces a transitional acceleration scale, constituting a much subtler phenomenon than that of the CDM halo, whose mass is of the order ($\sim$1000\%) relative to the disk mass.

To put the orders of magnitude into perspective from the outset (more detailed calculations can be found in Section \ref{sec3}), \(\dot M_d\) currently corresponds to a \(0.0000000067\%\) growth per year for a galaxy of mass \(M_d=10^{11}M_\odot\). This continuous growth may come about through various mechanisms, e.g., the steady accretion of IGM gas \citep{Sancisi2008GasAccretion}, the formation and evolution of cosmologically coupled black holes (e.g., \cite{Farrah2023BHCosmoCoupling,Croker2024DESI}), a fundamental relationship between gravitational binding energy and vacuum energy \citep{Fahr2012TheInfluence}, or a scale-invariant property of the gravitational potential \citep{10.1093/mnras/stad078}.

As for \(\dot r_d\), the greatest radial velocity of the entire disk corresponds to \(0.4\%\) of the smallest orbital velocity, well within the error bars of current measurements. For example, Hunt and Carlberg
\cite{Hunt_2016} estimate a solar rotation velocity around the Milky Way center of \mbox{\(v_{\odot}=239\pm9\) {km/s}}, and, for the case of a comparable galaxy (\(M_d=10^{11}M_{\odot}\)), the greatest radial velocity is \(\dot r_d=0.77\) {km/s}. Therefore, both \(\dot M_d\) and \(\dot r_d\) are seemingly negligible features of the present galactic dynamics, but their combination can explain the emergence of \(a_0\) from first principles.

Another crucial aspect to be highlighted at this point is the time dependence of the critical acceleration scale under CTEC conditions. Equation (2), expressed in terms of the present values \(a_0\) and \(H_0\), can be generalized in the following manner:

\begin{equation}
a_c(t)\approx\frac{cH(t)}{2\pi}
\end{equation}

Observational constraints on the time dependence of the critical acceleration scale are only starting to be established. The error bars are still far too big to allow any definitive conclusion \citep{10.1093/mnrasl/slae085,DELPOPOLO2024101414}, but in the near future, higher precision data will play a key role in this line of investigation. As the program marches on, it is essential to track the cosmology-dependent features of the data treatment, such as the age–redshift correlation (see \mbox{Section \ref{sec6}}), time dilation effects, flux corrections, etc. One must then adequately place each model in its respective cosmological background for a coherent analysis. Provided that such prerequisites are fulfilled, the measurement of galactic rotation curves at varying redshifts offers an unparalleled opportunity to test different models that explicitly predict \(a_c(z)\).

One such model is the Scale Invariant Vacuum (SIV), which posits that the macroscopic empty space is scale-invariant, homogeneous, and isotropic \citep{universe6030046}. Within SIV, since the mass of an object depends on the conformal scale factor \(\lambda\), varying like \(M(t)=M(t_0)(t/t_0)\), a similar type of Cosmological Time-Energy Coupling is present. Moreover, the radius of the object exhibits the same time scaling, resulting in a gravitational potential \(\Phi=GM/r\) that is fundamentally invariant and can play a similar role to the term \(G\dot M_d/\dot r_d\) discussed above. However, regarding the solution to the flat rotation curves in the absence of dark matter, the descriptions take different paths. In the SIV context, the rate of mass change is considered to be too small to have an appreciable effect for systems with timescales smaller than a few Myr. Only the conformal transformations of \(r\) and time itself with an approximately constant \(\lambda\) (\(r=\lambda r'\) and \(t=\lambda t'\)) are used to find a MOND-analog \mbox{behavior \citep{10.1093/mnras/stad078}}.

In our approach, on the other hand, the mass growth (although minor), coupled with the radius change, is precisely what gives rise to an acceleration scale that is proportional to the Hubble parameter. The result of these alternative perspectives is a pair of different predictions for \(a_c(z)\). Within SIV, the explicit formula is derived in \cite{10.1093/mnrasl/slae085}, and it clearly contrasts with the monotonic behavior herein demonstrated. This is an example of how observational progress can help us converge to the most predictive model.

\subsection{The Radial Acceleration Relation}

\begin{figure}
    \centering
    \includegraphics[width=1\linewidth]{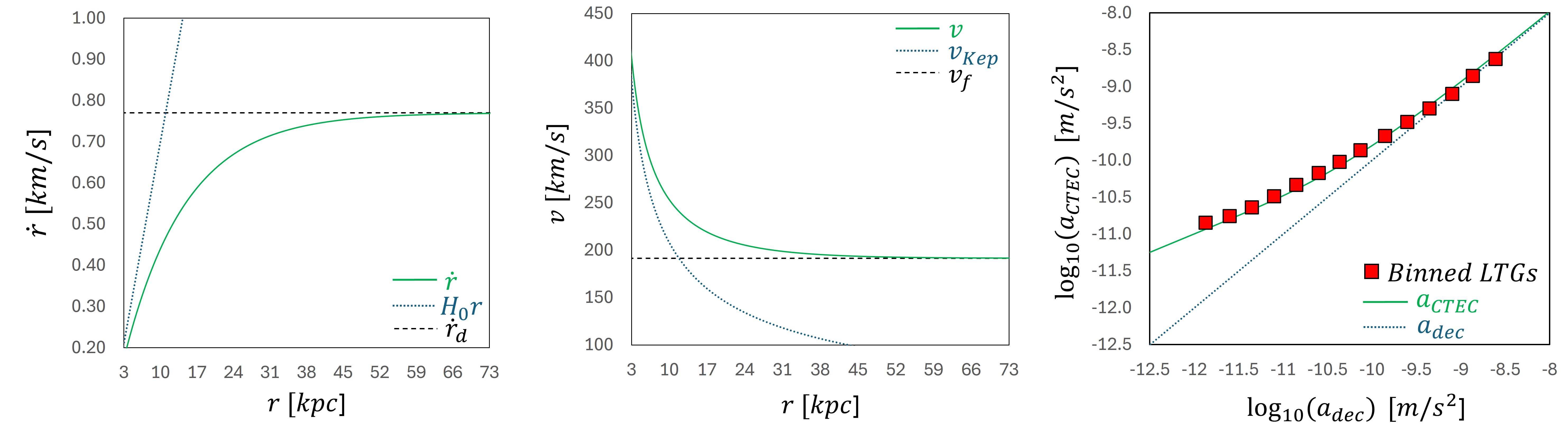}
    \caption{Transition from the expected Keplerian regime into the flat rotation regime for a bulge-dominated spiral galaxy of \(M_d=10^{11}M_{\odot}\). The left-hand side and central panels show, respectively, the radial and orbital velocities. The right-hand side panel shows the Radial Acceleration Relation (RAR) predicted by models that feature Cosmological Time-Energy Coupling (CTEC). The squares are the binned late-type galaxy data reported in \cite{Lelli_2017}.}
    \label{Figure1RAR}
\end{figure}

Thus far, we have focused on the temporal behavior of accelerations and speeds in the outer regions of the disk (\(r\approx r_d\)). Now, we turn our attention to the spatial dynamics of the spiral galaxies. The transition between the expected Keplerian behavior and the flat velocity regime can be visualized in Figure \ref{Figure1RAR}, and here, we work our way toward the radial acceleration relation.

To say that a galaxy is slowly growing over time implies the presence of outward radial velocities, which are significantly smaller than the orbital velocities (\(\dot r << v\)) at any given distance from the center so that the orbits are still approximately circular. Over the lifetime of the galaxy, the objects with greater radial velocities will have advanced to the most exterior regions of the disk (\(r\approx\dot rt_0\)), which can be written as \(\dot r \approx t_0^{-1} r\), tending to create the characteristic spatial profile \(\dot r\approx H_0 r\) inside the disk. 

Let us run our analysis outside the galactic bulge so that the mass does not vary significantly with the distance from the center any longer, i.e., \(M=M_d\) and \(\dot M=\dot M_d\), independently of \(r\). It is clear, then, that even for a growing galaxy, the inner orbital velocities are still inversely proportional to the square root of \(r\), displaying characteristic Keplerian behavior:

\begin{equation}
v_{Kep}=\sqrt{\frac{G\dot M}{\dot r}}=\sqrt{\frac{G\dot M}{H_0r}}
\end{equation}

However, the disk boundary condition requires \(\dot M/\dot r\) to approach the constant \(\dot M_d/\dot r_d\). Thus, a feedback loop starts: as the radial velocity \(\dot r\) flattens, the orbital velocity \mbox{\(v=(G\dot M/\dot r)^{1/2}\)} also flattens, which makes \(\dot r\) flatten more intensely still, and so on. The result is the exponential flattening of both velocities in the vicinities of the disk radius \(r_d\).

The curve that describes the radial velocities tracing a linear behavior \(\dot r\approx H_0 r\) inside the disk and exponentially approaching the asymptote \(\dot r_d\) in the outer regions of the galaxy is given by the equation

\begin{equation}
\dot r=\dot r_d\left(1-e^{-\frac{r}{r_d}}\right)
\end{equation}
yielding orbital velocities that initially follow Keplerian decline and then exponentially approach \(v_f\):

\begin{equation}
v=\sqrt{\frac{G\dot M}{\dot r}}=\sqrt{\frac{G\dot M_d}{\dot r_d\left(1-e^{-\frac{r}{r_d}}\right)}}=v_f\left(1-e^{-\frac{r}{r_d}}\right)^{-1/2}
\end{equation}

Finally, the observed radial acceleration resulting from the time-coupled growth of the galaxy is given by

\begin{equation}
a_{CTEC}=\frac{v^2}{r}=\frac{1}{r}\frac{G\dot M_d}{\dot r_d\left(1-e^{-\frac{r}{r_d}}\right)}=\frac{v_f^2}{r}\left(1-e^{-\frac{r}{r_d}}\right)^{-1}
\end{equation}
which can be confronted with the decoupled acceleration \(a_{dec}=GM_d/r^2\). Observation supports \(a_{CTEC}\), as evidenced by the right-hand side panel of Figure \ref{Figure1RAR}. The excellent agreement with the binned data reported by Lelli et al. \cite{Lelli_2017} can only be obtained when the approximation \(a_{cp}\approx a_0 \approx10^{-10}~\text{ms}^{-2}\) is valid in the vicinities of the disk radius. Naturally, no individual galaxy will have followed the evolution path given by Equations (13) exactly, but these time functions can statistically describe a typical population of spiral galaxies so as to successfully reflect the average radial acceleration relation at late times.

As an introduction to a novel mechanism for the emergence of \(a_0\), this work focuses on the simplest case of a bulge-dominated spiral galaxy. An extension to account for more complex shapes and mass distributions, which generate non-central forces and peculiar features on rotation curves (e.g., \citep{Sancisi_2004,Criss_Hofmeister_2020}), will follow.

\subsection{The Baryonic Tully-Fisher Relation}

We can now demonstrate that the existence of a critical acceleration that is an exclusive function of \(H_0\)---and independent of the specific mass and radius of the galaxy---is exactly what is needed to explain the Baryonic Tully-Fisher Relation (BTFR). This tight connection between the luminous content and orbital speeds is largely confirmed by observation, but it is difficult to fully reconcile with the CDM paradigm \citep{McGaugh_2000}.

The BTFR shows that the mass of the disk is proportional to the fourth power of its rotation curve’s asymptotic velocity.

\begin{equation}
M_d\propto v_f^4
\end{equation}

Again, by employing only Newtonian dynamics, one can demonstrate how such a relation may emerge. First, by isolating \(r\) in the centripetal acceleration equation

\begin{equation}
a_{cp}=\frac{v^2}{r}, \ \ \ r=\frac{v^2}{a_{cp}}
\end{equation}
and inserting this expression into the Newtonian formula for gravitational acceleration

\begin{equation}
a_{cp}=\frac{GM}{r^2}=\frac{GMa_{cp}^2}{v^4}, \ \ \ M=\frac{1}{Ga_{cp}}v^4
\end{equation}
it becomes evident that the BTFR will manifest itself from the point where the dominant centripetal acceleration is independent of any particular disk mass \(M\) and orbital velocity \(v\), so it can act as a constant in Equation (27). This is the exact role that \(a_0\) plays by being a sole function of \(H_0\) and not varying locally from galaxy to galaxy. Thus, in the inner part of the disk, the high accelerations will dictate the dynamics, but in the outer regions of the spirals, the transition around \(a_0\) occurs, accounting for the flattening of the rotation curves and the BTFR, which can now be expressed in terms of the fundamental cosmological parameters:

\begin{equation}
M_d=\frac{1}{Ga_0}v_f^4=\frac{2\pi}{GcH_0}v_f^4
\end{equation}

Furthermore, the BTFR defines the very radius of interest \(r_d\), which can be determined once \(v_f\) is measured:

\begin{equation}
r_d=\frac{1}{a_0}v_f^2=\frac{2\pi}{cH_0}v_f^2
\end{equation}

\section{The Galactic Pizza}\label{sec3}

In this section, a theoretical case study is explored. An imaginary spiral galaxy of mass \(M_d=10^{11}M_{\odot}\) is taken as a prototype for the gravity augmentation phenomenon (\(a_0\) emergence) via mass and radius growth.

Referring back to the evolution functions \(M_d=M_f+\dot M_dt\) and \(r_d=r_f+\dot r_dt\), suppose that, after an initial growth spurt resulting in \(M_f=8.5 \times10^9M_{\odot}=1.7\times10^{40}\) {kg} and \(r_f=1\) kpc \(=3.1\times10^{19}\) {m}, the galactic evolution is dominated by the rates of change \(\dot M_d=4.2\times10^{23}\) {kg/s} and \(\dot r_d=770\) {m/s}. After a Hubble time of \mbox{\(H_0^{-1}=14~\text{Gyr}=4.4\times10^{17}\) {s}}, the mass of the galaxy is, indeed, \(M_d=10^{11}M_{\odot}=2\times10^{41}\) {kg}, and its radius is \mbox{\(r_d=11.9\) {kpc} = \(3.7\times10\)$^{20}$ {m}}.

The values of \(M_f\), \(\dot M_d\), \(r_f\), and \(\dot r_d\) were chosen to comply with all the necessary conditions to make the peripheral orbital velocities flat and for the crossover acceleration to approach the observed order of magnitude \(a_0\sim 10^{-10}\) {ms}$^{-2}$.

\begin{equation}
\frac{M_f}{r_f}=\frac{\dot M_d}{\dot r_d}, \ \ \  r_d >>r_f, \ \ \ a_0=\frac{G\dot M_d}{\dot r_d^2}H_0=1.1\times10^{-10}~\text{m/s}^2
\end{equation}

Sure enough, when such conditions are met, the predicted radial acceleration relation shows excellent agreement with observation (see third panel of Figure \ref{Figure1RAR}). Naturally, there exist countless alternative combinations of \(M_f\), \(\dot M_d\), \(r_f\), and \(\dot r_d\) that would fulfill those criteria. All of such combinations produce satisfactory predictions for the RAR, while combinations that do not agree with at least one of the conditions above cannot generate RAR curves that reproduce the observed data.

It is also worth noting that the value of \(a_0\) depends on the chosen value of \(H_0\), which is here considered to be \(70\) {km s}$^{-1}${Mpc}$^{-1}$ or \(2.27\times10^{-18}\) {s}$^{-1}$. Therefore, the symmetric approach can be taken, and a high precision measurement of \(a_0\) for the spiral galaxies in the local universe may constitute an independent way of determining the Hubble constant \mbox{\(H_0=2\pi a_0/c\)}, an estimation technique of high relevance in the context of the “Hubble tension''~\citep{DiValentino2021HubbleTension}.

Now, in order to assess the impact of the spiral growth in our observations, it is important to have a second look at the rates of change \(\dot r_d\) and \(\dot M_d\). The allegorical nature of the following discussion is aligned with the purpose of conveying meaning to readers with both technical and non-technical backgrounds.

At first glance, the calculated values of radial velocity \(\dot r_d=770\) {m/s} and mass growth \(\dot M_d=4.23\times10^{23}\) {kg/s} may strike one as enormous quantities, far too great not to be immediately perceived upon observation of spiral nebulae. This is, however, a false impression resulting from our earthly perspective and lack of intuition for galactic scales. In order to resolve this issue, let us scale down our galaxy to the size of a family pizza (\(r_{pizza}=21\) {cm}) and watch this pizza grow over time with the same proportional speed as the galaxy in order to see how long it takes for any change to become apparent.

The proportional radius growth of the pizza would be \(\dot r_{pizza}=(r_{pizza}/r_d)\times \dot r_d\) = \((0.21/3.7\times10^{20})\times770=4.43\times10\)$^{-19}$ {m/s}, which can be converted to 70 {$\mu$m}/5 {Myr}. This means it would take 5 million years for the radius of the pizza to grow by the thickness of an average human hair. Likewise, if an advanced civilization manages to monitor this galaxy through a telescope for 5 million years, its radial increase would be exceedingly challenging to even notice. It is clear, then, that galactic growth is utterly imperceptible on a human timescale. As for the mass growth, considering that the family pizza weighs about \(1.5\) kg, its corresponding mass increase would be \(\dot M_{pizza}=(M_{pizza}/M_d)\times \dot M_d\) \mbox{= $(1.5/2\times10^{41})\times 4.23\times10^{23}=3.16\times10^{-18}$ {kg/s}}, which can be converted to 0.5 {g}/5 {Myr}. This means that, over the same 5 million years, the mass of our galactic pizza would grow by one-tenth of a teaspoon of mozzarella cheese.

The objective of this analogy is to express how tiny rates of change can yield the acceleration scale that flattens rotation curves and ultimately holds galaxies together. This is certainly an economical recipe for the emergence of \(a_0\), especially when confronted with the CDM paradigm, which would require an extra halo of 5 to 10 “dark pizzas'' around the visible disk to explain the same small acceleration.

\section{Cosmological Time-Energy Coupling}

The current understanding of cosmological evolution relies upon a continuous energy increase in the universe over cosmic time. In the \(\Lambda\)CDM framework, dark energy, in the form of a cosmological constant, is the mathematical entity that perpetually introduces new energy into the large-scale cosmos as it expands \citep{Carroll2001}. In such a paradigm, the total matter contribution for the energy budget of the universe (\(U_M\)), including both baryonic and dark matter, is fixed since the end of the re-heating epoch (\(U_M = k_1\)), yielding an energy density that decreases with volume expansion, i.e., with the cube of the scale factor (\(\rho_M\propto a^{-3}(t)\)). For radiation, the number of photons is fixed since the recombination era, but the radiation contribution to the energy content (\(U_R\)) decreases linearly with the cosmological redshift (\(U_R\propto a^{-1}(t)\)), resulting in an energy density that decreases with the fourth power of the scale factor (\(\rho_R\propto a^{-4} (t)\)). For the cosmological constant \(\Lambda\), however, the energy density is what is invariant (\(\rho_\Lambda =k_2\)), meaning that its total energy contribution must increase with volume expansion (\(U_\Lambda\propto a^3 (t)\)). The overall effect is a progressive growth of energy in the universe since early times, with an ever-increasing contribution from dark energy. The composite energy density (\(\rho=\rho_R+\rho_M+\rho_\Lambda\)), in turn, has always been decreasing, with the temporal scaling depending on the component that dominates each phase, though it is destined to be asymptotically constant in the future de Sitter regime (\(\rho\approx \rho_\Lambda\)).

With the observational developments of the past decades, starting with \cite{Riess1998Observational,Perlmutter_1999}, it is now difficult to envision a cosmological model that does not present some energy growth as an integral part of the expansion. There exist, however, descriptions that make the correlation between cosmic time and energy even more fundamental. While \(\Lambda\)CDM posits transitory phases, in which different components take turns to dominate the expansion rate and, therefore, the rate of energy accumulation, there are models that require energy to be a simple linear function of time. Here, such direct correlation is denominated “Cosmological Time-Energy Coupling''. The models that present CTEC combine two basic premises:

{(i) The} 
composite energy density of the cosmic fluid is consistently equal to the critical value throughout expansion history so that the geometry of the universe is always flat.

\begin{equation}
\rho=\rho_c=\frac{3c^2}{8\pi G}H^2
\end{equation}

{(ii) The} cosmic horizon is defined by the Hubble radius, which grows at the speed of light, as per Hubble's law \(c=H(t)R_H(t)\), yielding a fundamental relationship between the Hubble parameter and the cosmic age:
\begin{equation}
R_H=ct, \ \ \ H(t)=\frac{c}{R_H(t)}=\frac{1}{t}
\end{equation}

Hence, the resulting internal energy is linearly proportional to time:

\begin{equation}
U=\rho V \propto \frac{1}{t^2}\times t^3
\end{equation}

\begin{equation}
\therefore U \propto t
\end{equation}

It is worth noting that, counter-intuitively, the total conservation of energy is not violated when CTEC is present. The cumulative negative contribution of gravitational energy counterbalances the increment of positive internal energy in all spacetime, which can be readily seen by expressing the adiabatic case of the first law of thermodynamics:

\begin{equation}
\frac{dQ}{dt}= \frac{dU}{dt}+p \frac{dV}{dt}=0, \ \ \ \frac{dU}{dt}=-p \frac{dV}{dt}
\end{equation}

Gravity itself exerts a negative pressure as the volume increases with horizon expansion (\(V\propto R_H^3\)). Thus, one can determine the equation of state (EoS) \( \omega \triangleq p⁄\rho\) in the CTEC context:

\begin{equation}
\frac{d(\rho R_H^3)}{dt}=-p \frac{d(R_H^3)}{dt},
\end{equation}
which turns into the fluid or “continuity'' equation,

\begin{equation}
\frac{d\rho}{dt}+3H\rho(1+\omega)=0,
\end{equation}
whose single solution is \(\omega=-1/3\), given that \(\rho \propto t^{-2}\) and \(H= t^{-1}\).

Note that this is not the EoS of matter or any individual field on its own, but the EoS of the large-scale cosmic fluid that is expanding in the presence of gravity. In other words, gravity and the cosmic energy content are inherently coupled as a gravito-fluid. Furthermore, the overall energy balance of the universe is consistent with the maximally symmetric case; by weighing all the positive energy content on one side of the scale and the negative contribution from gravitational pressure on the other side of the scale, the result is always the Zero-Energy condition \(\rho+3p=0\).

This simple-looking computation is of profound significance. The very nature of gravity, with its negative energy contribution, allows for positive energy to grow in the universe over time, which is the same principle applied by Guth \citep{Guth1981PhysRevD.23.347} in the inflationary context. For some readers, a mass increase may also be reminiscent of the “Steady-State Universe'' defended by Hoyle \citep{Hoyle1948MNRAS.108..372H}. Note, however, that the cosmological time-energy coupling herein studied still results in an evolving (cooling) cosmos with decreasing energy density and, \mbox{therefore, temperature.}

The outcome of this analysis is that “CTEC'' can be a double acronym, standing for “Cosmological Time-Energy Coupling'' as well as “Cosmological Total Energy Conservation''.

From Section \ref{sec:Newtoninan}, it becomes evident why the dark matter puzzle is naturally resolved within the CTEC context, at least on galactic scales. After all, what is needed for \(a_0\) to emerge is a systematic increase in mass over time and an expansion regime. Thus, the gravitationally bound structures (in this case, spiral galaxies) are fundamentally connected to and actively participate in the universal behavior of internal energy growth over time. Indeed, this ought to be expected if the CTEC realizes itself in our universe, i.e., if the cosmic mass-energy increases monotonically, the structures are bound to exhibit this systematic increase as well, as they are focal points of energy concentration in the universe. The fact that this connection explains the effects associated with dark matter on galactic scales only elevates the explanatory power of CTEC models, even though this fundamental coupling was not conceived to fulfill such a purpose. Rather, it is the behavior of a flat universe that possesses the critical energy density throughout its expansion history and whose age is always the reciprocal of the Hubble parameter. The same solution is not necessarily irreconcilable with other cosmological models, such as \(\Lambda\)CDM, but it would be harder to apply, demanding additional assumptions that are, as of yet, unknown.

On the topic of the cosmological expansion's influence on local systems, investigations date back to the early days of relativistic cosmology \citep{McVittie}. Yet, to this day, there is no general or definitive solution. Rather, advances have been made in particular cases, i.e., specific matter arrangements embedded in pre-determined metrics. One such \mbox{development \citep{Price_2012}} revealed the “all-or-nothing'' behavior based on a classical atom bound by electrical attraction in a de Sitter space: if the coupling is weak, the atom comoves with the expanding universe; if the coupling is strong, the atom completely ignores its cosmological background after an initial transient perturbation. At face value, if the interpretation of this result is generalized, it appears that strongly bound gravitational systems resist cosmological expansion and are only marginally perturbed. However, the authors of \cite{PhysRevD.76.063510} concluded that the assumption of a de Sitter background is too limiting for a generalization to be assumed, especially in a universe that is not a de Sitter space. They show that a strongly gravitating central object embedded in an FLRW background can, in fact, comove with the expansion.

Although general solutions are still lacking and approximations debated for even some of the most classical and studied systems (see also \cite{GIULINI201424,Spengler_2022} for reviews), the interplay between local and cosmological dynamics is of great interest to this work and will be the object of future analyses. In particular, we aim to understand the mechanisms through which a system may exchange energy with the cosmic fluid, whose equation of state is \(\omega=p/\rho=-1/3\), in three alternative backgrounds: flat FLRW, inertial structure expansion in a Minkowskian metric, and conformal scale-invariant vacuum. To this end, approximations must be closely monitored and controlled. As studied in Section \ref{sec2.2}, even a \(0.0067\% \) growth per {Myr} can give rise to a critical acceleration scale; hence, subtle effects cannot be promptly neglected.

\section{The Cosmological Acceleration Scale}

Thus far, we have mathematically traced back the origin of the ubiquitous acceleration scale \(a_0\sim 10^{-10}\) {ms}$^{-2}$ to the growth rates of galactic masses and radii. From this starting point, we can now expand the analysis. Whenever discussing a physical parameter, it is essential to look at it from multiple angles if possible. Fortunately, the link between \(a_0\) and the Hubble constant \(H_0\) grants us the opportunity to carry out this very deed. Since the time-coupled growth proposition for structures has a cosmological counterpart (models positing that \(\rho=\rho_c\) and \(H t=1\) must undergo mass-energy growth over time), it is only natural that the examination of \(a_0\) from a cosmic perspective should yield fruitful outcomes.

Starting with Equations (2) and (31), one can express the Hubble constant in \mbox{two convenient ways:}

\begin{equation}
H_0=\left(\frac{8 \pi G \rho_c}{3c^2} \right)^{1/2}, \ \ \ H_0 = \frac{2 \pi}{c}a_0,
\end{equation}
from which it becomes clear that the acceleration scale \(a_0\) is fundamentally related to the current critical density of the universe:

\begin{equation}
a_0=\left(\frac{2G\rho_c}{3\pi}\right)^{1/2}
\end{equation}

In fact, in the CTEC context, the acceleration scale can be expressed in various forms in terms of \(H_0\) and \(t_0\) and the current values of \(R_H\) and \(\rho_c\):

\begin{equation}
a_0= \frac{cH_0}{2\pi}=\frac{c}{2\pi t_0}= \frac{c^2}{2\pi R_H}=\frac{4}{3}\frac{G\rho_c}{cH_0}= \frac{4}{3}\frac{G\rho_ct_0}{c}= \frac{4}{3}\frac{G\rho_cR_H}{c^2}
\end{equation}

Most interestingly, one can express \(a_0\) in a familiar field form if the current mass of the entire universe \(M_H\) (contained within the Hubble radius \(R_H\)) enters the analysis:

\begin{equation}
\rho_c=\frac{M_Hc^2}{\frac{4\pi}{3}R_H^3}
\end{equation}

By substituting \(\rho_c\) in the last expression of (40), we obtain

\begin{equation}
a_0=\frac{1}{\pi}\frac{GM_H}{R_H^2}
\end{equation}

The conclusion one can derive is the following: the acceleration scale that flattens the rotation curves of spiral galaxies can be expressed as a function of the gravitational influence of the observable universe on the luminous disk. This creates a tight connection between the development of galaxies and the evolution of the large-scale cosmos. 

Furthermore, \(M_H\) and \(R_H\) can be obtained directly from the fundamental cosmological parameters. The expressions below are applicable to any time, but here, we include their current values for reference:
\begin{equation}
M_H=\frac{\rho_c}{c^2}\frac{4\pi}{3}R_H^3=\frac{c^3}{2G}t_0=9\times10^{52}~\text{{kg}}, \ \ \ R_H=\frac{c}{H_0}=ct_0 = 1.32\times10^{26}~\text{{m}}
\end{equation}
and their rates of change are:
\vspace{-6pt}
\begin{equation}
\dot M_H=\frac{c^3}{2G}=2\times10^{35}~\text{{kg/s}}, \ \ \ \dot R_H=c=3\times10^8~\text{{m/s}}
\end{equation}




Note that \(\dot M_H\) is time-invariant, constituting a fixed combination of \(c\) and \(G\). Expanding this analysis within the realm of fundamental constants, one can relate this growth rate to the Planck units of mass (\(m_p\)), time (\(t_p\)), and length (\(l_p\)):

\begin{equation}
m_p=\sqrt{\frac{\hbar c}{G}}=2.2\times10^{-8}~\text{{kg}}, \ \ \ t_p=\sqrt{\frac{\hbar G}{c^5}}=5.4\times10^{-44}~\text{{s}}, \ \ \ l_p=\sqrt{\frac{\hbar G}{c^3}}=1.6\times10^{-35}~\text{{m}}
\end{equation}

\begin{equation}
\therefore \dot M_H=\frac{c^3}{2G}=\frac{1}{2}\frac{m_p}{t_p}
\end{equation}

Therefore, the cosmological time-energy coupling also assigns an important significance to the Planck mass: it manifests on greater scales relative to other Planck units because it is of cosmological relevance. It represents the order of magnitude in which mass grows in the entire universe over each tiny unit of Planck time.

Now, to better understand the role of the factor \((\frac{1}{2})\), it is relevant to discuss the coupling between the Hubble radius and the Hubble mass. Starting from Equation (42) and expressing \(a_0\) as \(c^2/(2\pi R_H)\), it becomes evident that

\begin{equation}
R_H=\frac{2GM_H}{c^2}
\end{equation}

This is a familiar relation given that, for any generic mass \(m\), the corresponding Schwarzschild radius (\(r_s\)) is defined as
\begin{equation}
r_S = \frac{2Gm}{c^2}
\end{equation}

Therefore, the Hubble radius is exactly the Schwarzschild radius corresponding to the Hubble mass. Now, relating the Planck length and mass
\begin{equation}
l_p=\frac{Gm_p}{c^2}
\end{equation}
and expressing this in the form of a Schwarzschild radius
\begin{equation}
l_p=\frac{2G(\frac{1}{2}m_p)}{c^2},
\end{equation}
it becomes clear that a Schwarzschildian relation \(r_s(m)\) is maintained when, for each radius change of one \(l_p\), a corresponding change of \((\frac{1}{2})m_p\) takes place. Hence, the cosmological growth scaling is the following: for each Planck time, \(R_H\) grows by a Planck length, and \(M_H\) grows by half a Planck mass. This yields a growing \(R_H\) that always corresponds to the Schwarzschild radius of the observable universe.

This equivalence between the cosmological and Schwarzschild horizons may be a hint that the evolution of the universe is fundamentally linked to the evolution of black holes. A two-way analysis can be established: our universe itself may be a black hole immersed in a greater cosmos (see, for example, \citep{Pathria72,SMOLIN2004705,sym14091849}), and a natural logical leap would imply that black holes in our universe may systematically evolve like nested universes, being cosmologically coupled (see additionally \cite{Farrah2023BHCosmoCoupling,Croker2024DESI}). On its own, the Schwarzschildian relation observed from within does not constitute definitive proof of such a fundamental link, but it certainly points towards a fascinating possibility in the CTEC context, which will be addressed in future collaborations.

\subsection*{Notes on Total Energy Budget and Dark Energy}
Although the scope of this work focuses on the galactic problem, in this section, we briefly discuss how the CTEC defines a new picture for the cosmic energy budget in order to secure internal consistency. It may seem paradoxical at first to equate the universal energy density to the critical value and remove dark matter within the same framework. After all, this extension to the standard model should be responsible for about \(27\%\) of the energy budget that composes the critical density. To understand how one can reconcile the CTEC conditions with the absence of this contribution, let us run a spatial analysis at the present epoch: the observable universe in the \(\Lambda\)CDM paradigm has a current radius of ~46.5~{Gly}, whereas models that feature the CTEC properties have the expected observable radius of ~14 {Gly}. Note that, in the context of \(\Lambda\)CDM, multiple horizons emerge (particle horizon, event horizon, and Hubble sphere), and the “observable universe'' is taken to be defined by the particle horizon \citep{Davis_Lineweaver_2004}. On the other hand, for CTEC-type models, no superluminal recession takes place, and the cosmological horizon is delimited by the surface where the recession speed equals \(c\). Thus, the Hubble sphere defines the “observable universe''. Such difference in radius translates into an even more expressive volumetric ratio, meaning that the energy needed to fill the CTEC-type universes and obtain the same critical density is only \((14/46.5)^3=2.7\%\) of the mass-energy content required by the \(\Lambda\)CDM model. Therefore, the baryonic matter alone is sufficient to fill the current CTEC volume and still yield the critical density. In fact, this result may aid in the discussions regarding the “Missing Baryon Problem'', for it has been a long-term challenge to even detect all the \(5\%\) prescribed by \(\Lambda\)CDM \citep{McGaugh_2007}. Only recently have new studies analyzing the IGM claimed to get closer \citep{Yang2022}, but by interpreting and extrapolating the results according to the \(\Lambda\)CDM scales. All the data analysis must be reviewed from the CTEC perspective in the future since, in this description, energy content increases over time, so it is expected that the total mass decreases with the redshift term \((1+z)\).

Most importantly, as the baryonic matter may single-handedly account for all the critical energy density, no cosmological constant should be needed to make the universe flat. In a coherent manner, when \(t=H^{-1}\), no acceleration occurs, and the dimming of standard candles observed in \cite{Riess1998Observational,Perlmutter_1999} can be reproduced by CTEC-type models with good accuracy \citep{Chodorowski2005Cosmology,Melia2019}. Therefore, a cosmological time-energy coupling in the expanding cosmic fluid can, in principle, account for the observables associated with both components of the dark sector.

\section{Early Galaxy Evolution}\label{sec6}

Two elements must be considered in the attempt to elucidate the observation of massive structures at high redshifts, i.e., to explain the most recent data concerning \( M(z)\). The first one is the very conversion between the redshift itself and cosmic age, or \(t(z)\), which is model-dependent. The second one is the rate of matter agglomeration towards potential wells in the early epochs, or \(M(t)\), and whether or not a dark component as the source of augmented gravity is necessary and sufficient. In short, to explain \(M(z)\), one must understand \(t(z)\) and \(M(t)\).

\begin{figure}

\centering 
\includegraphics[width=1.\linewidth]{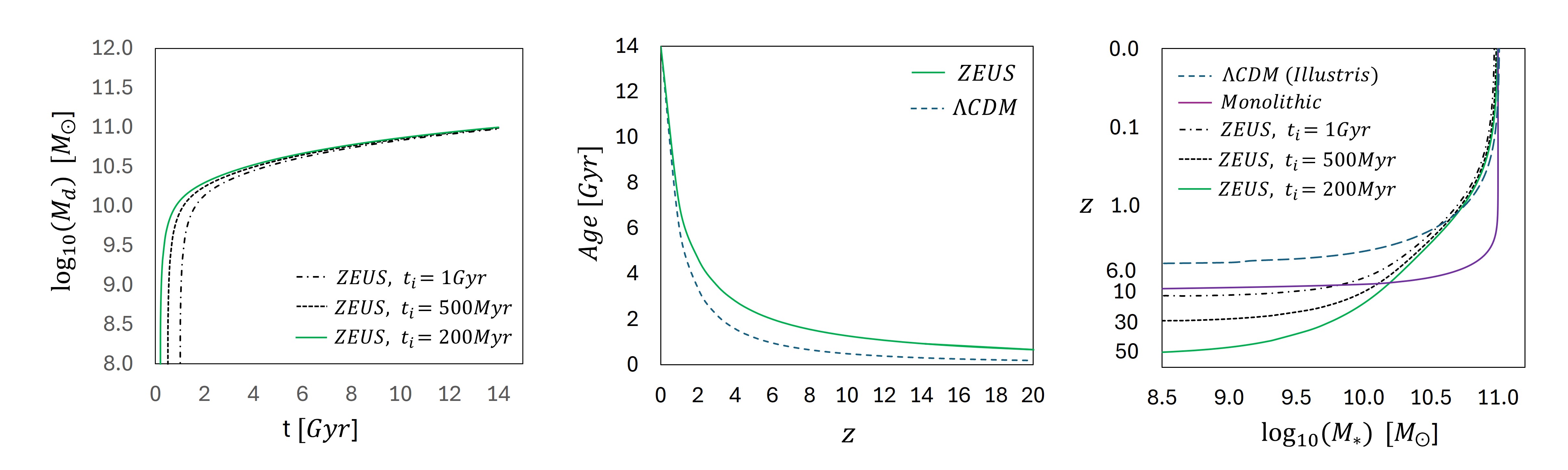}

\caption{{Galaxy} 
mass evolution according to ZEUS, a CTEC-type model. The left-hand side panel shows \(M(t)\), as per Equation (13). For all curves \(M_f=8.5\times 10^{9}M_{\odot}\), \(\tau_1=500\) {Myr},
and \(\dot M_d=6.7M_{\odot}/\) {yr}. The only discerning parameter is \(t_i\): for the solid green curve, it takes 200 {Myr} for evolution to start after the big bang, changing to 500 {Myr} for the dotted line and 1 {Gyr} for the dash-dotted line. The central panel shows \(t(z)\) for the vanilla \(\Lambda\)CDM model (dashed blue line) with \(\Omega_{\Lambda}=0.7\), \(\Omega_{M}=0.3\), and \(H_0=70\) {kms}$^{-1}${Mpc}$^{-1}$, compared to \(t(z)\) for the ZEUS model (solid green line). The right-hand side panel shows \(M_*(z)\), normalized to \(M_*(0)=10^{11}M_{\odot}\), for different functions of galaxy evolution and background cosmologies. The dashed blue line is the median curve from the Illustris simulation \citep{Illustris2016}, the solid purple line is the monolithic process, as presented in \citep{McGaugh_2024}, and the three ZEUS curves follow the same logic as the first panel.}

\label{figure2}
\end{figure}

In terms of \(t(z)\), CTEC-type models can immediately alleviate the “impossibly early galaxy problem'' \citep{Steinhardt2016TheImpossibly} since they assign a greater cosmic age to a given redshift when compared to the \(\Lambda\)CDM model, especially in the early universe. The central panel of \mbox{Figure \ref{figure2}}, extracted from \cite{Novais2024}, shows that the ZEUS model, which fulfills the CTEC conditions, decompresses the evolution time by a factor of \(2-4\) at \(6\lesssim z\lesssim20\). To give an example, the recently confirmed record-breaker JADES-GS-z14-0, observed at \(z=14.32\) \citep{Carniani2024JADES}, had \mbox{290 {Myr}} to evolve according to \(\Lambda\)CDM but 914 {Myr} in the ZEUS timeline.

The simple \(t(z)\) function inherent to CTEC models and applicable to both ZEUS and \(R_h=ct\) \citep{Melia2014PrematureFormation} is

\begin{equation}
t_{source}=\frac{t_0}{(1+z)}=\frac{14~\text{Gyr}}{(1+z)}
\end{equation}

In the ZEUS model, the cosmic fluid expands in the spatially flat Minkowskian coordinates. In this metric, the total time dilation perceived by an observer receiving signals from a receding source is the combination of the Lorentzian factor \(\gamma=1/\sqrt{1-(v/c)^2}\) due to the velocity itself and an extra factor \((1+\beta)=(1+v/c)\) due to the continuous increase in separation. This yields a total time dilation factor \(\gamma(1+\beta)=\sqrt{(1+\beta)/(1-\beta)}\) that is identical to the relativistic Doppler redshift term \((1+z)\).

\begin{equation}
(1+z)=\frac{\sqrt{1+\frac{v}{c}}}{\sqrt{1-\frac{v}{c}}}\times \frac{\sqrt{1+\frac{v}{c}}}{\sqrt{1+\frac{v}{c}}}=\frac{1+\frac{v}{c}}{\sqrt{1-\left(\frac{v}{c}\right)^2}}=\gamma(1+\beta)=\frac{t_0}{t_{source}}
\end{equation}

Thus, the age conversion is given by Equation (51) (more details in \cite{Novais2024,Davis_Lineweaver_2004,Chodorowski2005Cosmology}).

In the \(R_h=ct\) model, the FLRW scale factor is a linear function of time (\(a(t)= t/t_0\)) and also the reciprocal of \((1+z)\), directly yielding Equation (51).

Turning to \(M(t)\), the temporal evolution of cosmic structures must be determined, at least in statistical terms. As discussed in Section \ref{sec2.2}, the standard hierarchical framework is built upon the idea that cold dark matter is necessary to explain the early potential wells towards which the first progenitors coalesce. Subsequently, consecutive mergers would be responsible for the advent of the present large structures. Observations at redshifts \(z<6\) were used to calibrate simulations, and a great merit of the hierarchical picture is its ability to generate a good resemblance to actual measurements in this redshift range \citep{Illustris2016}.

However, as observations from HST and JWST gather evidence for highly massive objects at far greater redshifts, the previously calibrated CDM parameters cease to be a sufficient explanation. McGaugh et al. \cite{McGaugh_2024} show that such discoveries at \(z>6\) point to a monolithic-like process, though recognizing that the exact mechanism for a practically instantaneous structure formation at \(z\sim 10\) is still lacking.

In this work, we propose a mass evolution function that reconciles a first growth spurt with progressive buildup over cosmic time. We showed that such an evolution can explain the augmented gravity on galactic scales observed in the local universe. Moreover, the time dependence of the critical acceleration scale means that an even greater degree of “extra gravity'' was present in earlier times, facilitating rapid structure formation.

Naturally, simulations must also be utilized in the future to confirm the viability of such mechanisms in a coherent cosmological background, i.e., under CTEC conditions. However, as a first step, one can directly apply the proposed evolution function and compare it to the hierarchical and monolithic processes. In order to use the same basis that the authors of \cite{McGaugh_2024} show in their Figure 1, we consider the stellar mass function to roughly follow the total mass function:

\begin{equation}
M_*(t)=M_*^f\left(1-e^{\frac{-(t-t_i)}{\tau_1}}\right)+\dot M_*t,
\end{equation}
and then, on the third panel of Figure \ref{figure2}, we plot the evolution of a galaxy with present stellar mass of \(10^{11}M_{\odot}\).

In general, the resulting ZEUS curves show notorious agreement with the CDM Illustris project precisely in the region where the latter works best with observations, i.e., \(z<6\). For higher redshifts, the ZEUS functions deviate from the CDM curve---greatly due to the different \(t(z)\)---and account for the observation of galaxies in advanced evolution stages (\(M_*\sim10^9-10^{10}M_{\odot}\)) at \(z\gtrsim 10\). At the lower limit (\(z\sim10\)), this advantage is shared with the monolithic curve.

The only parameter that differentiates the three ZEUS curves is \(t_i\), or the time it takes for evolution to start after the big bang. As we continue to observe at progressively higher redshifts, we should be able to constrain this term in the near future.

As an initial step, the evolution function herein proposed not only explains the flat rotation curves on galactic scales at late times but also goes a long way towards reconciling---at earlier times---the conventional hierarchical formation curve with the empirical monolithic approximation, bringing together their respective advantages at lower and higher redshifts. Simulation efforts are required to advance this study.

\section{Conclusions}

The puzzle of augmented gravity in galaxies carries an interesting dichotomy. If approached from a pure acceleration perspective, it can be considered a subtle problem since the discrepancies only start to manifest themselves around a tiny acceleration scale (\(a_0\sim 10^{-10}\) {ms}$^{-2}$). However, if considered through a missing-mass lens, the size of the problem appears to escalate copiously to the extent that one has yet to detect \(85\%\) of the universe’s mass, uncovering its exotic nature along the way. Though it appears disproportional, if one assumes we are solving the gravitational equations correctly, postulating and pursuing dark matter is the reasonable path to take. Then again, faced with null results, a parallel question may be asked: are we even solving the correct equations? Thus, the exercise of modifying gravity also becomes relevant. An upside of this investigation path is the return to the small acceleration discrepancy realm. However, modifying gravity in specific regimes without a more fundamental theory or underlying phenomenon does not come without its problems, and the dichotomy remains present both in the theoretical and observational arenas.

Without a definitive solution, one may return to the question of whether we are solving the equations correctly and, more specifically, if we are exhausting all the possibilities they contain. Pondering upon Newtonian gravity on an as-is basis may seem an obsolete endeavor, but sometimes, one must think deeply about simple things. By allowing minor rates of change for mass and radius in the Newtonian formulation, an acceleration scale can be obtained, which naturally conforms to the critical value extracted from observation. As explored in this work, the requisite changes are of tenuous order against the broader dynamics of the spiral galaxies, and yet they have the power to explain the small acceleration discrepancies from an underlying phenomenon that is cosmologically motivated: a subtle solution to a subtle problem.

\bmhead{Acknowledgments}
The authors are grateful to the anonymous referees for their detailed and constructive reviews, which have led to significant improvements in the content and presentation of the manuscript.






\end{document}